\documentclass[aps,prl, twocolumn, floatfix,superscriptaddress,showpacs,showkeys]{revtex4-1}
\usepackage{epsf,amsmath,amssymb,verbatim,color,multirow,pifont,graphicx,cleveref,tabularx,cancel,soul}
\usepackage[dvipsnames]{xcolor}
\usepackage{rotating}

\setstcolor{red}

\crefname{equation}{Eq.}{Eqs.}
\crefname{section}{Sec.}{Secs.}
\crefname{figure}{Fig.}{Figs.}
\crefname{table}{Table}{Tables.}

\newcommand{\nubar}{\bar \nu}

\newcommand{\tcm}{t_c^{-}}
\newcommand{\tcp}{t_c^{+}}

\begin{document}

\title{Critical phenomena of a hybrid phase transition in cluster merging dynamics}
\author{K. Choi}
\affiliation{CCSS, CTP and Department of Physics and Astronomy, Seoul National University, Seoul 08826, Korea}
\author{Deokjae Lee}
\affiliation{CCSS, CTP and Department of Physics and Astronomy, Seoul National University, Seoul 08826, Korea}
\author{Y. S. Cho}
\affiliation{Department of Physics, Chonbuk National University, Jeonju 54896, Korea}
\author{J. C. Thiele}
\affiliation{Computational Physics for Engineering Materials, Institute for Building Materials, ETH Z\"urich, 8093 Z\"urich, Switzerland}
\author{H. J. Herrmann}
\affiliation{Computational Physics for Engineering Materials, Institute for Building Materials, ETH Z\"urich, 8093 Z\"urich, Switzerland}
\author{B. Kahng}
\email{bkahng@snu.ac.kr}
\affiliation{CCSS, CTP and Department of Physics and Astronomy, Seoul National University, Seoul 08826, Korea}
\date{\today}

\begin{abstract}
Recently, a hybrid percolation transitions (HPT) that exhibits both a discontinuous transition and critical behavior at the same transition point has been observed in diverse complex systems. In spite of considerable effort to develop the theory of HPT, it is still incomplete, particularly when the transition is induced by cluster merging dynamics. Here, we aim to develop a theoretical framework of the HPT induced by such dynamics. We find that two correlation-length exponents are necessary for characterizing the giant cluster and finite clusters, respectively. Finite-size scaling method for the HPT is also introduced. The conventional formula of the fractal dimension in terms of the critical exponents is not valid. Neither the giant nor finite clusters are fractals but they have fractal boundaries.

\end{abstract}

\maketitle


Percolation has long served as a simple model that undergoes a geometrical phase transition in non-equilibrium disordered systems~\cite{stauffer}. As an occupation probability $p$ is increased beyond a transition point $p_c$, a macroscopic-scale giant cluster emerges across the system. Theory of percolation transition was well established  by the Kasteleyn-Fortuin formula~\cite{fortuin}. This percolation theory has been used for understanding percolation-related diverse phenomena such as conductor--insulator transitions~\cite{con_insul}, the resilience of systems~\cite{resilience1,resilience2,resilience3}, the formation of public opinion~\cite{opinion1,opinion2}, and the spread of disease in a population~\cite{disease,review_epidemics}.  The percolation transition is known to be one of the most robust continuous transitions~\cite{stauffer,review_epjb}. 

Recently, however, many abrupt percolation transitions have been observed in complex systems~\cite{science_achilioptas,riordan,cho_science,souza_nphy,review_jstat,enclave_nphy,enclave_prl}, for instance, large-scale blackouts in power grid systems~\cite{mcc1} and pandemics~\cite{grassberger_nphy}, in which the order parameter increases abruptly at a transition point.  Among those transitions, an HPT has attracted substantial attention. The transitions in $k$-core percolation~\cite{kcore1,kcore2,kcore3,kcore4,kcore5} and in the cascading failure model on interdependent networks \cite{mcc1,mcc2,mcc3,mcc4,mcc5} are prototypical instances of the HPT. 
For these cases, the HPT is driven by cascade failures over the entire system as links are removed.
The cluster size distribution (CSD) does not obey a power law. Instead, the avalanche size distribution follows a power law and shows critical behavior~\cite{mcc5}. Consequently, the conventional formalism of percolation transition based on the CSD cannot be extended in an appropriate way to the HPT. 


Here, we aim to develop a theoretical framework of the critical phenomena of the HPT. To achieve this goal, we use a modified version~\cite{r_ER_hybrid} of the so-called half-restricted percolation model~\cite{half} in two and infinite dimensions. This model has potential applications to the transport or communication systems with global control equipments~\cite{r_ER_hybrid}. This model exhibits a HPT induced by cluster merging dynamics as links are added. The order parameter remains zero up to a transition point, at which it increases abruptly to a finite value, leading to a first-order transition. As the order parameter abruptly increases, clusters are self-organized according to size: the size distribution of finite clusters obeys a power law, and the giant cluster is located separately from the finite clusters. Next, the order parameter increases continuously and exhibits a second-order phase transition. We show that indeed the properties of the second-order transition can be determined by the power-law behavior of the CSD. However, we need two correlation-length exponents, $\nu_g$ and $\nu_s$, to characterize the formation of the giant cluster in finite systems and the size of finite clusters, respectively. To obtain those results, we extended the finite-size scaling method, which is useful to explore HPTs.  

We first recall the theory of continuous percolation transitions~\cite{stauffer}. The order parameter increases from zero continuously as $m\sim (p-p_c)^{\beta}$ for $p > p_c$; the mean cluster size diverges as $\langle s \rangle \sim |p-p_c|^{-\gamma}$, and the correlation length diverges as $\xi \sim |p-p_c|^{-\nu}$. These exponents are related by the hyperscaling relation $2\beta+\gamma=d\nu$, where $d$ is the spatial dimension. Those critical exponents and the scaling relations can be obtained from the CSD denoted as $n_s(p)$, which behaves as $n_s(p)\sim s^{-\tau}\exp(-s/s^*)$, where $s^*\sim (p-p_c)^{-1/\sigma}$. On the other hand, a percolating cluster at a transition point $p_c$ is a fractal object. The total number of sites belonging to this percolating cluster, denoted as $M_{\infty}(L)$, where $L$ is the linear size of the system in the Euclidean space, scales as $M_{\infty}(L)\sim L^{D}$, where $D$ is the fractal dimension. In addition to the percolating cluster, finite clusters also behave similarly, as $M_s(R_s)\sim R_s^D$, where $R_s$ is the linear size of a finite cluster of size $s$. The fractal dimension $D$ is related to the critical exponents $\beta$ and $\nu$ as $D=d-\beta/\nu$.

We introduce the so-called restricted percolation model on a square lattice of size $L\times L$. $N=L^2$ is the total number of sites in the system. Occupation of bonds during percolation is achieved dynamically~\cite{ben_naim}. Bonds are added to the system one by one at each time step according to a given rule. When $\ell$ bonds are added to the system, the occupation probability $p$ conventionally used in bond percolation corresponds to $p=\ell/(2L^2)$. Here we use a control parameter $t=\ell/N$, which is equivalent to $t=2p$. The dynamic rule of bond occupation is as follows: At each time step, we classify clusters into two sets, a set $R$ and its complement set $R^c$ according to their sizes. The set $R$ contains the smallest clusters, those satisfying $\sum_{i = 1}^{k - 1} s(c_i) < \lfloor gN \rfloor \le \sum_{i = 1}^{k} s(c_i)$, where $s(c_i)$ denotes the size of the cluster with index $c_i$, and $g \in [0,1]$ is a parameter that controls the size of set $R$. Then set $R$ contains the smallest $k$ clusters $\{c_1,c_2, \cdots, c_k\}$, and the set $R^c$ contains the remaining large clusters. Next, we occupy a randomly chosen unoccupied bond, one or both ends of which belong to the clusters in $R$. We do not allow the occupation of bonds between two sites belonging to clusters in $R^c$.    

Initially, there is no occupied bond, and each node is an isolated cluster. Then the set $R$ is composed of $gN$ randomly selected nodes. As the occupation dynamics continues, growth of large clusters is suppressed because the occupation of bonds between two clusters belonging to $R^c$ is not allowed. When all the clusters belong to set $R$, the dynamic rule becomes equivalent to that of ordinary bond percolation. On the other hand, if $g = 1$, the process is exactly equivalent to ordinary bond percolation from the beginning. We use periodic boundary conditions in the simulations. We call this model the restricted percolation model with reference to the original name, the half-restricted percolation model~\cite{half}. We remark that our dynamic rule is slightly different from the original one in that a cluster on the boundary between the two sets in the original model is regarded as an element of set $R$ in our model. 

\begin{figure}[]
\centering
\includegraphics[width=1.0\linewidth]{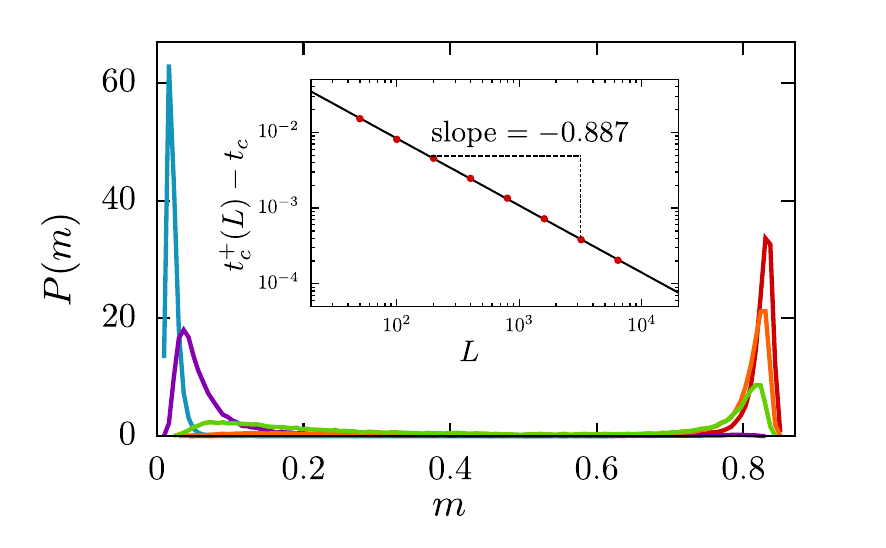}
\caption{ Plot of the distributions of the order parameter for $L = 6400$. The distribution is unimodal with a peak near zero (left peak) for $t < \tcm(L)$ (light blue and violet). As $t$ passes $\tcm(L)$ (light green), the right peak begins to grow, whereas the left-hand peak shrinks. As $t$ reaches $\tcp(L)$ (orange), the left peak disappears and the distribution becomes unimodal with the right peak alone. The two characteristic times $\tcm(L)$ and $\tcp(L)$ converge to a transition point $t_c$ in the thermodynamic limit. Inset shows the scaling behavior $\tcp(L)-t_c\sim L^{-1/\nu_g^{\prime}}$.}
\label{fig:fig1}
\end{figure}

Determining a transition point $t_c$ is not straightforward in the HPT as we used the conventional finite-size scaling method in a second-order percolation transition. To proceed, we first characterize two time steps, $\tcm(L)$ and $\tcp(L)$, for finite systems of linear size $L$ using the distribution of the order parameter $P(m; t, L)$ obtained from different configurations for fixed $t$ and $L$. The order parameter is defined as $m(t)=M_{\infty}(t)/N$. For $t \le \tcm(L)$, $P(m; t, L)$ exhibits a peak at a certain $m$ near $m=0$ denoted as $m^-(L)$. $m^-(L)$ is denoted as $m_*^-(L)$  at a particular point $t=t_c^-(L)$, which satisfies the criterion that for $t > t_c^-(L)$, $P(m)$ begins to exhibit another peak at $m^+(L)$ near $m=1$  as shown in Fig.~\ref{fig:fig1}. As $t$ is increased further, the peak at $m^-$ shrinks, whereas the other peak at $m^+$ becomes higher. At $t = \tcp(L)$, the peak at $m^-$ disappears, and only the peak at $m_*^+$ remains. For $t > \tcp(L)$, a peak remains at $m^+(L)$, which grows with $t$. As $L$ is increased, $\tcm(L)$ and $\tcp(L)$ converge to a certain value $t_c$, and $m^-(L)\to 0$, and $m^+(L)$ approaches a certain value $m_0$. This suggests that the order parameter exhibits a discontinuous jump at $t_c$ in the thermodynamic limit.
Particularly, we find that $\tcp(L)-t_c\sim L^{-1/\nu_g^{\prime}}$, in which the exponent $\nu_g^{\prime}$ is estimated to be  $\nu_g^{\prime}\approx 1.13\pm 0.07$ for $g=0.5$.

\begin{figure}[]
\centering
\includegraphics[width=1.0\linewidth]{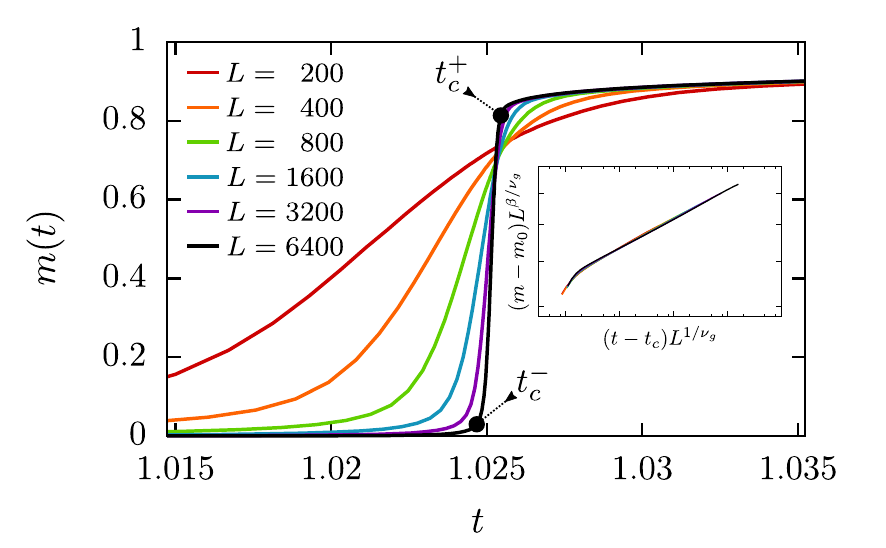}
\caption{Plot of the order parameter $m(t)$ versus $t$ for the restricted percolation model with $g=0.5$ for different lateral sizes $L$ and in data collapse form of $(m-m_0)L^{\beta/\nu_g}$ versus $(t-t_c)L^{1/\nu_g}$ (inset). Characteristic time steps $\tcm$ and $\tcp$ for $L=6400$ are marked.}
\label{fig2}
\end{figure}

In finite systems, the order parameter $m(t)$ is approximately $m^-(L)$ for $t \le  \tcm(L)$, but it increases rapidly in the interval $\tcm(L) < t < \tcp(L)$, and it becomes $m_*^+(L)$ at $\tcp(L)$, beyond which it increases gradually as $t$ is increased as shown in Fig.~\ref{fig2}. In the thermodynamic limit, $m(t)$ behaves as  
\begin{equation}\label{eq:op}
m(t)=\left\{
\begin{array}{lr}
0 & ~{\rm for}~~  t < t_c, \\
m_0+r(t-t_c)^{\beta} & ~{\rm for}~~ t \ge t_c, 
\end{array}
\right.
\end{equation}
where $m_0$ and $r$ are constants. $m_0$ represents the fraction of sites belonging to the giant cluster at $t_c$, and the second term represents the increment of the giant cluster size divided by $N$ as $t$ is increased beyond $t_c$. In finite systems, the order parameter for $ t > t_c$ may be written as $m(t)-m_0 \sim L^{-\beta/\nu_g}$ in the critical region above $t_c$, in which the lateral size of the system $L$ is less than the correlation length of the giant cluster, $\xi_g \sim (t-t_c)^{-\nu_g}$. We determine the critical exponent $\beta$ to be $\beta=0.061\pm 0.005$ for $g=0.5$ by plotting $m-m_0$ versus $t-t_c$, while we find the exponent $\nu_g$ to be $\nu_g=1.03\pm 0.08$ for $g=0.5$ by scaling plotting $(m-m_0)L^{\beta/\nu_g}$ versus $(t-t_c)L^{1/\nu_g}$ for different system sizes (see Fig.~\ref{fig2}). Because the numerical values of $\nu_g$ and $\nu_g^{\prime}$ agree within the error bars, we may regard them as being the same. We examine the susceptibility in the form of the fluctuations of the order parameter. We obtain the associated exponent as $\gamma_m=1.79\pm 0.08$ for $g=0.5$ (see the SM). The scaling relation $2\beta+\gamma_m=d\nu_g$ is  satisfied within error bars.

\begin{figure}[]
\centering
\includegraphics[width=1.0\linewidth]{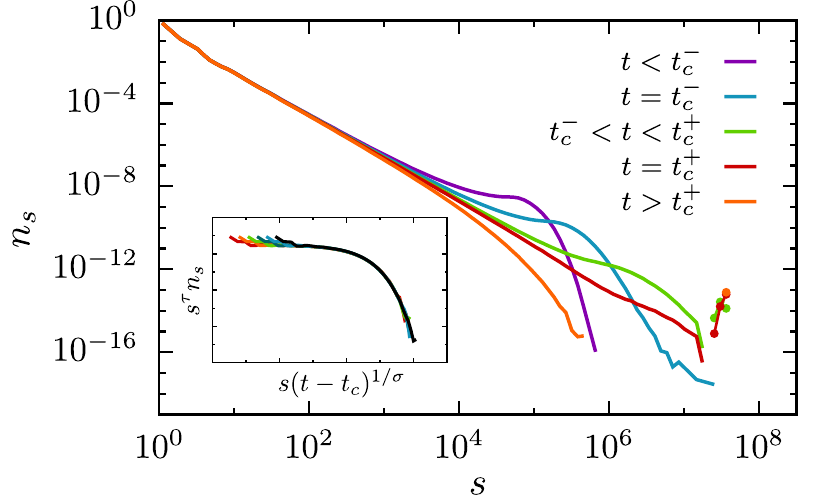}
\caption{Plot of the size distribution of finite clusters $n_s(t)$ and the giant cluster (separated dots) at various time steps for $L = 6400$ and $g=0.5$. At $t < \tcp(L)$, a bump exists in the tail part, but it shrinks as $t$ increases, and it finally disappears at $t =\tcp(L)$. Inset: scaling plot of the size distribution of finite clusters in the form of $s^{\tau}n_s$ versus $s(t-t_c)^{1/\sigma}$ for several $t > t_c$.} 
\label{fig3}
\end{figure}

The size distribution $n_s(t)$ of finite clusters exhibits a power-law decay with the exponent $\tau$ at a transition point $t_c$. This is an important feature of the critical behavior of the HPT at $t_c$. In finite systems, the power-law behavior occurs at $t_c^+(L)$ (see Fig.~\ref{fig3}), which is reduced to $t_c$ as $L \to \infty$. When $t > t_c$, the size distribution of finite clusters exhibits crossover behavior at $s^*$: It undergoes a power-law decay for $s < s^*$ but an exponential decay for $s > s^*$. Thus, $n_s(t)\sim s^{-\tau}e^{-s/s^*}$, where $s^*\sim (t-t_c)^{-1/\sigma}$. We obtain the exponents $\tau$ and $\sigma$ from Fig.~\ref{fig3}. Using the results of $\tau=2.035\pm 0.009$ and $\sigma=0.58\pm 0.03$ for $g=0.5$ and the scaling relation $\beta=(\tau-2)/\sigma$, we determine an alternative value of the critical exponent $\beta$ as $0.0613\pm 0.0187$ for $g=0.5$. This value is consistent with the directly measured one within error bar. We examine the susceptibility in the form of the second moment of the size distribution of finite clusters, and obtain the associated exponent as $\gamma=1.56\pm 0.15$. The scaling relation $\gamma=(3-\tau)/\sigma$ is satisfied within error bar (see the SM).

In finite systems, the cluster size has a finite cutoff $s^*$ resulting from the finite-size effect. Introducing the correlation length $\xi_s$ of finite clusters as $\xi_s \sim (t-t_c)^{-\nu_s}$ and when $L \le \xi_s$, we obtain  that $s^*\sim L^{1/\sigma\nu_s}$. We numerically obtain that $1/(\sigma \nu_s) \approx 1.9523\pm 0.0045$ for $g=0.5$ by measuring the ratio of the third moment of $n_s(t)$ to the second moment. Similar values are obtained for the ratio of the $(n+1)$-th moment to the $n$-th moment, where $n \ge 3$. We also obtain $\sigma=0.58\pm 0.03$ for $g=0.5$ from the size distribution of finite clusters. Thus, the exponent $\nu_s$ is expected to be $\nu_s=0.886 \pm 0.048$ for $g=0.5$. On the other hand, the exponent $\nu_s$ can be obtained using the scaling relation $\nu_s=(\tau-1)/\sigma d$. Using the directly measured values $\tau=2.035\pm 0009$ and $\sigma=0.58\pm 0.03$, we obtain $\nu_s=0.895 \pm 0.054$, which is consistent with the value obtained above. Thus, the hyperscaling relation $(\tau-1)/\sigma=d\nu_s$ is satisfied. 

We also check the presence of the two exponents $\nu_g$ and $\nu_s$ in the mean-field limit~\cite{r_ER_hybrid}. In previous studies, we found numerically that $\beta=0.21\pm 0.05$, $\gamma=0.83\pm 0.05$, $1/\sigma=1.04\pm 0.05$, and $\tau=2.18\pm 0.04$ for $g=0.5$. These exponent values yield ${\bar \nu_s}\equiv d\nu_s=(\tau-1)/\sigma \approx 1.23\pm 0.10$ for finite clusters. On the other hand, we obtain ${\bar \nu_g}\equiv  d\nu_g\approx 2.0$ using the data collapse technique for the formula $(m(t)-m_0)N^{\beta/{\bar \nu_g}}$ versus $(t-t_c)N^{1/{\bar \nu_g}}$ for different system sizes. This numerical result is presented in the supplemental material (see the SM). Therefore, the two  exponents, ${\bar \nu_g}$ and ${\bar \nu_s}$, are different even in the mean-field version of the restricted percolation model. 

Based on those finite-size scalings for the giant cluster and finite clusters, we confirm that we need two exponents, $\nu_g$ and $\nu_s$, associated with giant cluster and finite clusters, respectively, and that they are different.

When we take the ensemble average of a physical quantity numerically in the interval $\tcm(L) < t < \tcp(L)$, we need to note the following. For a given $t$, a giant cluster may have already formed in some realizations, although it may not in others. We need to separate the two cases when we take the average of a certain quantity over different realizations.  In this interval, however, the distribution $P_m$ is broad and the two peaks may not be pronouced (the green curve in Fig.~\ref{fig:fig1}), which means that it is impractical to separate the two cases. In contrast, the majority of realizations have no giant cluster in $t < t_c^-(L)$ and have a giant cluster in $t > t_c^+(L)$. Thus the ensemble average taken over the realizations in those separate regions can be easily calculated. Moreover the finite size effect is still observed near $t_c^-(L)$ and $t_c^+(L)$. Thus we use the simulation data obtained only in $t < t_c^-(L)$ or $t > t_c^+(L)$ for finite size scaling analysis and discard the data obtained in $t_c^-(L) < t < t_c^+(L)$. The asymptotic behavior of the system at $t_c^+(L)$ as $L \to \infty$ gives the behavior of the system as $t$ approaches $t_c$ from above in the thermodynamic limit, i.e., the properties of the percolating phase near the critical point. Similarly the properties of the non-percolating phase near the critical point are obtained by observing the system at $t_c^-(L)$. For instance, the scaling plot of $(m-m_0)L^{\beta/\nu_g}$ versus $(t-t_c)L^{1/\nu_g}$ was drawn in the region $t \ge t_c^+(L)$.

 Next, we are interested in the fractal dimensions of the giant and finite clusters. Here we determine these fractal dimensions using the box-covering method as follows: For a given cluster, we determine its center of mass. Then we open a window of size $\ell \times \ell$, the center of which is placed at the center of mass of the cluster. We count the number of occupied sites within the window, which is called the mass, $M(\ell)$. We obtain the average mass of the giant cluster, $M_{\infty}(\ell)$, over different configurations and the average mass of finite clusters of size $s$, $M_s(\ell)$,  over different clusters of the same size $s$ and different configurations. We also calculate the mean radius of gyration of all clusters of size $s$, denoted as $R_s$. 

\begin{figure}
\centering
\includegraphics[width=0.9\linewidth]{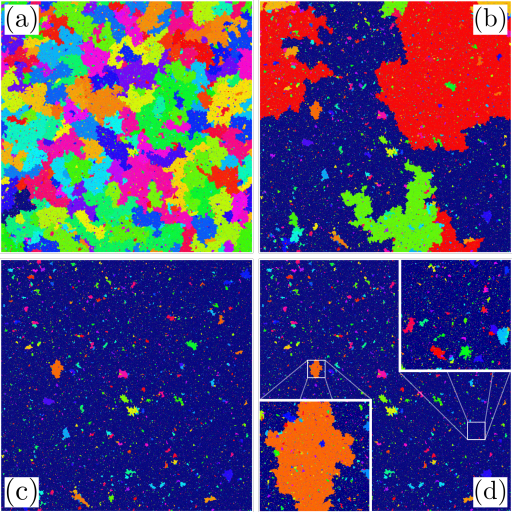}
\caption{Snapshots of the system of lateral size $L=6400$ with $g=0.5$ at three time steps (a) $t=\tcm \approx 1.02457346$, (b) $t_c \approx 1.02523$ and (c) $\tcp \approx 1.02543437$. At $\tcp $, the size distribution of finite clusters follows a power law. (d) Zoom-in snapshots of the giant cluster (top-right) and a finite cluster (lower-left)}
\label{fig:snapshots}
\end{figure}

We measure the fractal dimension $D_g$ of the giant cluster using the relation $M_{\infty}(\ell)\sim \ell^{D_g}$ for each system size $L$ at a transition point $t_c^+(L)$. As shown in Fig.~\ref{fig:snapshots}, the clusters are almost compact, and we obtain that $D_g=2.0001\pm 0.003$ regardless of $L$. We measure the fractal dimension of finite clusters using the relation $M_s(\ell)/s \sim (\ell/R_s)^{D_s}$. We obtain that $D_s=1.993\pm 0.010$ independent of the system size $L$. Therefore, we conclude that neither the giant cluster nor finite clusters are fractal in hybrid percolation. The conventional formalisms of the fractal dimension, $D=d-\beta/\nu$ and $D=1/\sigma \nu$, are not valid for the HPT.     
We remark that the transition point $t_c^+(L)$ of the HPT is larger than that of the ordinary bond percolation model, $t_c=1$. Thus, the number of occupied bonds in the critical region of the HPT is as dense as that in the supercritical region of ordinary percolation. Accordingly, the giant cluster as well as the finite clusters are almost compact with dimension $D_g=D_s=2$.  On the other hand, we examine the fractal property of the perimeter of the largest cluster at $t_c$. Using the yardstick method, we find that the accessible boundaries of the compact clusters at $t_c$ are fractal with a dimension less than $4/3$ for $g < 1$. We speculate the fractal dimension of the boundary to be the same as $D_b\approx 1.217\pm 0.001$, the fractal dimensions of the watershed~\cite{watershed} and the Gaussian model for the explosive percolation~\cite{hans}.

In summary, we investigated the critical phenomena of an HPT induced by cluster merging dynamics using the restricted percolation model. We showed that the characteristic sizes of the giant cluster and finite clusters scale separately using the exponents $\nu_g$ and $\nu_s$, respectively. The hyperscaling relations $2\beta+\gamma_m=d\nu_g$ and $(\tau-1)/\sigma=d\nu_s$ hold, respectively. They are different for the HPT, but the same for the ordinary percolation. These results are valid for any $g < 1$, even though individual critical exponents values depend on $g$. Numerical values of those exponents for different $g$ are listed in the SM. We found that the conventional relationship between the critical exponents and the fractal dimension of the giant and finite clusters breaks down. Moreover, the area of the giant and finite clusters are almost compact with the dimension $D_g=D_s=2$. However, the boundary of the giant cluster is fractal. Finally, we remark that the finite scaling analysis for HPTs is technically complicated. We introduced a new finite-size scaling method to determine the critical exponents in finite systems. We anticipate the finite-size scaling method and theoretical scheme to be useful for further exploration of HPTs.

\begin{acknowledgments}
This work was supported by the National Research Foundation of Korea by Grant No. NRF-2014R1A3A2069005. HJH thanks the European Research Council (ERC) Advanced Grant No. 319968-FlowCCS for financial support.
K.C. and D.L. contributed equally to this work.
\end{acknowledgments}

\clearpage\onecolumngrid

\appendix
\section*{Supplemental Material for Critical phenomena of a hybrid phase transition in cluster merging dynamics}

In this supplemental material, we first present numerical results for the susceptibilities of finite clusters and the giant cluster for the restricted percolation model in two dimensions. Second, we present the scaling plot of the order parameter versus $\Delta t = t-t_c$ for different system sizes in the mean-field limit for the restricted percolation model. Finally, we present the list of the critical exponents for the restricted percolation model with various values of $g$ in two dimensions.   

\subsection{Susceptibilities of the restricted percolation model in two dimensions}
\begin{figure}[h]
\centering
\includegraphics[width=0.5\linewidth]{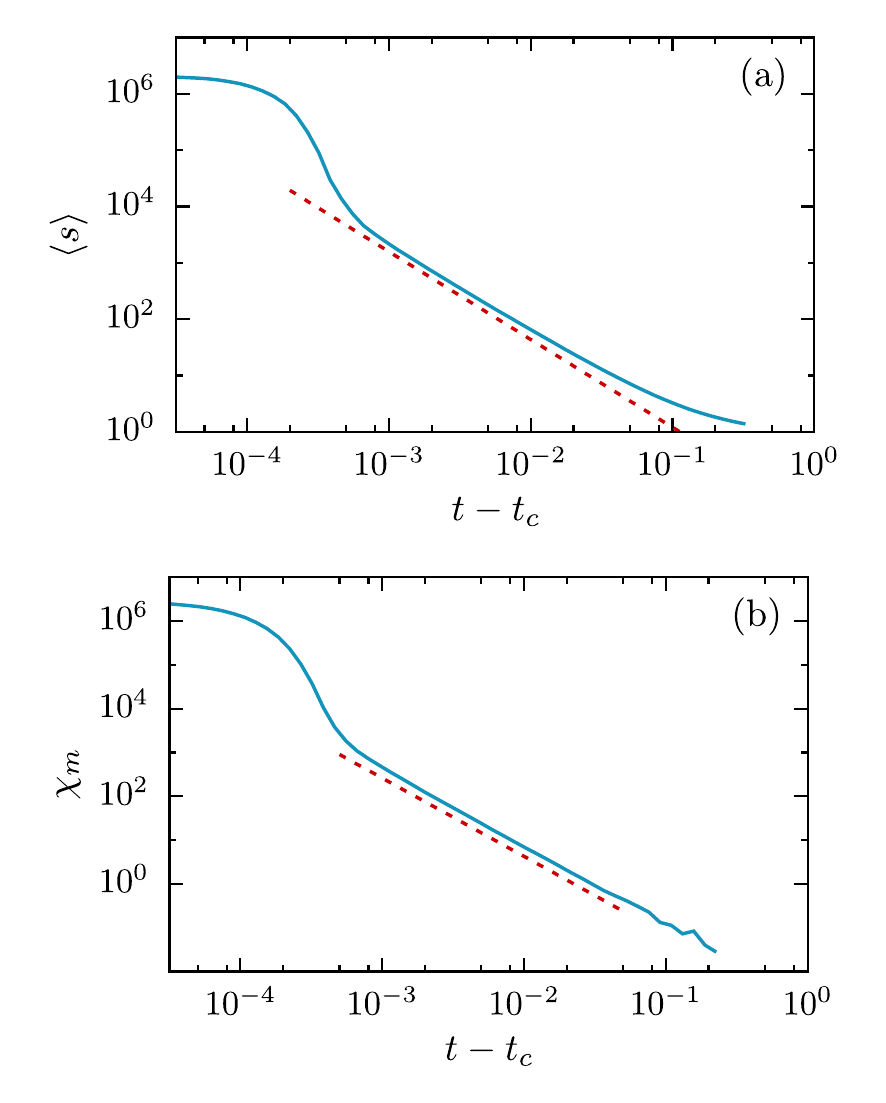}
\caption{(a) Plot of the susceptibility defined as $\chi_s=\langle s \rangle\sim \sum_{s=1}^{\rm finite} s^2 n_s$ versus $t-t_c$. $\chi_s \sim (t-t_c)^{-\gamma}$ is expected. Dashed line is a guideline with slope $-1.56$. The susceptibility exponent is obtained as $\gamma=1.56\pm 0.15$. (b) Plot of the susceptibility defined as $\chi_m=L^2(\langle m^2 \rangle -\langle m \rangle^2)$ versus $t-t_c$. $\chi_m\sim (t-t_c)^{-\gamma_m}$ is expected. Dashed line is a guideline with slope $-1.79$. The susceptibility exponent is obtained as $\gamma_m=1.79\pm 0.08$. Simulation data are obtained from the systems with lateral size $L=6400$.}
\label{smfig1}
\end{figure}
\newpage
\subsection{The restricted Erd\H{o}s-R\'enyi percolation model}
\begin{figure}[h]
\centering
\includegraphics[width=0.5\linewidth]{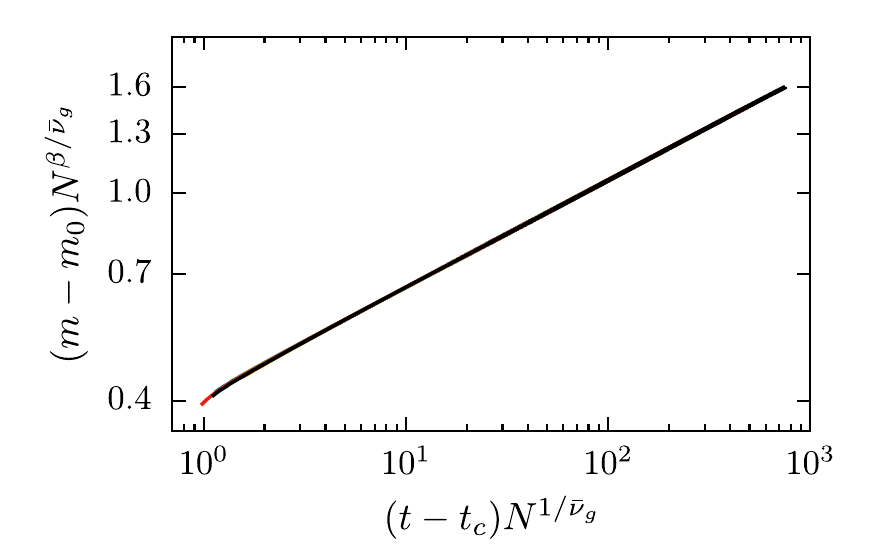}
\caption{{\bf Finite-size scaling of the order parameter for the restricted Erd\H{o}s-R\'enyi model:} Plot of the order parameter $m(t)$ versus $t$ for the restricted ER percolation model with $g=0.5$ for different system sizes $N$ in data collapse form of $(m-m_0)N^{\beta/\nubar_g}$ versus $(t-t_c)N^{1/\nubar_g}$. $\nubar_g$ is estimated to be $\approx 2.0$.
The system sizes are chosen as $N=2^0\times 10^4$ and $2^{12}\times 10^4$.}
\label{smfig2}
\end{figure}

\subsection{List of the critical exponents for the restricted percolation model with general $g$ in two dimensions}
\bgroup
\def\arraystretch{1.5}
\begin{sidewaystable}
\centering
\caption{The critical exponents for the restricted percolation model with general $g$ in two dimensions}
\setlength{\tabcolsep}{0.7em}
\begin{tabular}{ccccccccccc}
\hline
$g$
& $t_{c}$
& $m_{0}$
& $\beta$
& $\gamma_{m}$
& $\tau$
& $\sigma$
& $\gamma$
& $\nu_{g}^{\prime}$
& $\nu_{g}$
& $\nu_{s}$
\\
\hline\hline
0.1 
&$	1.06381	\pm	0.00004	$
&$	0.98	\pm	0.05	$
&$	0.003	\pm 0.005	$
&$	2.07	\pm	0.25	$
&$	1.918	\pm 0.035	$
&$	0.61	\pm 0.04	$
&$	1.77	\pm 0.25	$
&$	1.06	\pm	0.07	$
&$	0.95	\pm 0.12	$
&$	0.848	\pm	0.069	$
\\0.2
&$	1.06019	\pm	0.00004	$
&$	0.88	\pm 0.05	$
&$	0.016	\pm	0.008	$
&$	1.69	\pm 0.15	$
&$	1.993	\pm	0.015	$
&$	0.60	\pm 0.03	$
&$	1.56	\pm 1.15	$
&$	1.07	\pm	0.07	$
&$	0.95	\pm 0.10	$
&$	0.848	\pm	0.046	$
\\0.3
&$	1.05124	\pm	0.00004	$
&$	0.80	\pm	0.05	$
&$	0.038	\pm 0.008	$
&$	1.68	\pm 0.08	$
&$	2.023	\pm	0.015	$
&$	0.60	\pm 0.03	$
&$	1.53	\pm 0.15	$
&$	1.07	\pm	0.07	$
&$	0.99	\pm 0.10	$
&$	0.840	\pm	0.048	$
\\0.4 
&$	1.03903	\pm	0.00005	$
&$	0.68	\pm	0.05	$
&$	0.051	\pm 0.005	$
&$	1.73	\pm	0.08	$
&$	2.030	\pm 0.010	$
&$	0.58	\pm 0.03	$
&$	1.55	\pm 0.15	$
&$	1.09	\pm	0.05	$
&$	1.02	\pm 0.08	$
&$	0.883	\pm	0.050	$
\\0.5 
&$	1.02523	\pm	0.00005	$
&$	0.55	\pm	0.03	$
&$	0.061	\pm 0.005	$
&$	1.79	\pm 0.08	$
&$	2.035	\pm	0.009	$
&$	0.58	\pm 0.03	$
&$	1.56	\pm 0.15	$
&$	1.13	\pm	0.07	$
&$	1.03	\pm 0.08	$
&$	0.886	\pm	0.048	$
\\0.6
&$	1.01151	\pm	0.00005	$
&$	0.49	\pm	0.05	$
&$	0.086	\pm 0.005	$
&$	1.89	\pm	0.08	$
&$	2.047	\pm 0.009	$
&$	0.55	\pm 0.03	$
&$	1.64	\pm 0.15	$
&$	1.14	\pm	0.07	$
&$	1.09	\pm 0.08	$
&$	0.929	\pm	0.052	$
\\0.7
&$	0.99855	\pm	0.00006	$
&$	0.36	\pm	0.05	$
&$	0.098	\pm 0.005	$
&$	2.01	\pm	0.08	$
&$	2.050	\pm 0.009	$
&$	0.51	\pm 0.03	$
&$	1.77	\pm 0.15	$
&$	1.14	\pm	0.07	$
&$	1.11	\pm 0.09	$
&$	1.009	\pm	0.061	$
\\0.8
&$	0.98775	\pm	0.00006	$
&$	0.23	\pm	0.05	$
&$	0.108	\pm 0.005	$
&$	2.08	\pm	0.08	$
&$	2.052	\pm 0.008	$
&$	0.49	\pm 0.03	$
&$	1.79	\pm 0.15	$
&$	1.13	\pm	0.07	$
&$	1.15	\pm 0.08	$
&$	1.062	\pm	0.066	$
\\0.9
&$	0.98167	\pm	0.00010	$
&$	0.10	\pm	0.05	$
&$	0.120	\pm 0.008	$
&$	2.24	\pm	0.08	$
&$	2.053	\pm 0.009	$
&$	0.44	\pm 0.03	$
&$	2.03	\pm 0.17	$
&$	1.11	\pm	0.11	$
&$	1.20	\pm 0.09	$
&$	1.196	\pm	0.083	$
\\1.0
&$	1	$
&$	0	$
&$	5/36(0.139)	$
&	-
&$	187/91(2.055)	$
&$	36/91(0.40)	$
&$	43/18(2.39)	$
&
&$	4/3(1.33)	$
&
\\
\hline
\end{tabular}

\end{sidewaystable}
\egroup

\end{document}